\documentclass[11pt]{article} 

\usepackage{amsmath,amsthm,latexsym,amssymb,amsfonts,epsfig}

\newtheorem{Proposition}{Proposition}[section]

\newcommand{\be}{\begin{equation}}
\newcommand{\ee}{\end{equation}}
\newcommand{\ba}{\begin{eqnarray}}
\newcommand{\ea}{\end{eqnarray}}

\title{{\sf Smooth, invariant orthonormal basis}\\
 {\sf for singular potential Schr\"odinger operators}} 
\author{
{\sf J. Neuser}$^1$\thanks{{\sf 
jonas.neuser@gravity.fau.de}},
{\sf T. Thiemann}$^1$\thanks{{\sf 
thomas.thiemann@gravity.fau.de}}\\
\\
{\sf $^1$ Inst. for Quantum Gravity, FAU Erlangen -- N\"urnberg,}\\
{\sf Staudtstr. 7, 91058 Erlangen, Germany}\\
}
\date{{\small\sf \today}}

\makeatletter
\@addtoreset{equation}{section}
\makeatother

\begin{document} 

\maketitle

{\sf

\begin{abstract}
In a recent contribution we showed that there exists a smooth, dense 
domain for singular potential Schr\"odinger operators on the real line 
which is invariant 
under taking derivatives of arbitrary order and under multiplication by 
positive and negative integer powers of the coordinate. Moreover, inner 
products between basis elements of that domain were shown to be easily 
computable analytically.    

A task left open was to construct an orthonormal basis from elements of that 
domain by using Gram-Schmidt orthonormalisation. We perform that step 
in the present manuscript. We also consider the application of these methods 
to the positive real line for which one can no longer perform the integrals 
analytically but for which one can give tight analytical estimates.
\end{abstract}

\section{Introduction}
\label{s1}

In a previous paper we have communicated the observation that the span 
of the
functions on the real line  $x\mapsto \;b_n(x):= 
x^{-n}\; e^{-[x^2+x^{-2}]}$ with 
$n\in \mathbb{Z}$ not only forms a dense invariant domain for Schr\"odinger 
type Hamiltonians but also inner products between such functions can 
be computed analytically in closed form for which an explicit formula 
was provided. 

Important open questions included 1. whether 
a Gram-Schmidt orthonormalisation of the system can be provided in closed 
form and 2. how to proceed when the real axis is replaced by the positive 
real axis. In this paper we answer question 1 affirmatively and with regard 
to question 2 we provide elementary but tight estimates which are analytically
computable. We also relate the inner product formula provided in 
\cite{1} to the Bessel functions of the second kind. 
\\
The architecture of the present paper is as follows;\\
\\

In section \label{s2} we perform an explicit Gram-Schmidt orthonormalisation
of the $b_n$. The construction is actually not via the inner product 
formula provided in \cite{1} but uses the completeness relation of 
the Hermite functions $x\mapsto e_n(x),\;n\in \mathbb{N}_0$ on the real line. 
The explicit ONB formed from the $b_n$ is then unsurprisingly 
closely related to the Hermite functions: It is given by the functions 
$g_{j,n}(x)=x^{-j}\; e_n(x-x^{-1}),\;j=0,1;\; n\in \mathbb{Z}$. Thus far 
away from the origin these are just Hermite functions times $x^j$. That 
we need two families $n\mapsto g_{0,n}, \; g_{1,n}$ rather than just 
one is due to the desire to have an invariant domain for both multiplication
operators by $x, x^{-1}$ respectively. The action of the operators 
$d/dx, x, 1/x$ on that basis is explicitly provided and is conveniently 
expresssed in terms of annihilation and creation operators with 
respect to $z:=x-1/x$.

In section \ref{s3} we provide the estimates for inner products on the 
positive real axis for which the methods of \cite{1} are used. We give 
also an alternative method of computation.

In section \ref{s4} we summarise and conclude. In particular, we 
relate the inner product formula of \cite{1} to Bessel functions.

\section{Orthonormal basis and completeness relation}
\label{s2}

The notation is copied from paper \cite{1} where the smooth, dense, invariant 
domain defined by the linear span or the functions 
$b_n(x)=x^n\; e^{-a[x^2+x^{-2}]/2}$ was introduced where $a$ is a positive 
constant. The inner product of these functions in the Hilbert space 
${\cal H}=L_2(\mathbb{R},dx)$ can be computed explicitly, they form 
a basis but not an orthonormal basis. \\
\\
Consider the function
\be \label{2.1}
f:\; \mathbb{R}\to \mathbb{R};\; x\mapsto\; f(x):=x-x^{-1}
\ee
This is a double cover of the real axis and consequently the inverse of $f$
has two branches when solving $z=f(x)$ for $x$
\be \label{2.2}
x=\left\{ \begin{array}{cc} 
h_+(z):=\frac{z}{2}+\sqrt{1+[\frac{z}{2}]^2} & x\ge 0\\
h_-(z):=\frac{z}{2}-\sqrt{1+[\frac{z}{2}]^2} & x\le 0
\end{array}
\right.
\ee
The functions $h_\pm$ are strictly monotonously increasing 
since $|z/2|<w(z):=\sqrt{1+[z/2]^2}$ and satisfy the identities
\be \label{2.3}
h_+(z)\; h_-(z)=-1,\; dh_\pm/dz=\pm\;\frac{1}{2w}\;h_\pm  
\ee
Accordingly we have 
\be \label{2.4}
h_\pm(z)=h_\pm(z')\;\;\Leftrightarrow\;\; z=z'
\ee
To obtain an orthonormal basis in the linear span of the $b_n$ for $a=1$
we proceed 
as follows. Given $x,x'>0$ we find unique $z,z'\in \mathbb{R}$ such that
$x=h_+(z),x'=h_+(z')$. Using the well known distributional identity
\be \label{2.5}
\delta(F(z))=\sum_{F(z_0)=0}\; \frac{\delta(z-z_0)}{|dF/dz(z_0)|}
\ee
applied to $F(z)=h_+(z)-h_+(z')$ we have 
\ba \label{2.6}
\delta(x-x') &=& \delta(h_+(z)-h_+(z'))=
\frac{\delta(z-z')}{|dh_+/dz(z')|}
\nonumber\\
&=& \delta(z-z')\;\frac{2w(z')}{|h_+(z')|}
=\delta(z-z')\;\frac{2w(z')}{h_+(z')}
=\delta(z-z')\;\frac{h_+(z')-h_-(z')}{h_+(z')}
\nonumber\\
&=& \delta(z-z')\;[1+\frac{1}{h^2_+(z')}]
= \delta(z-z')\;[1+\frac{1}{h_+(z)\;h_+(z')}]
\ea
where we used (\ref{2.3}). Let 
\be \label{2.7}
e_n(z)=\frac{[A^\ast]^n}{\sqrt{n!}}\Omega,\;
A=2^{-1/2}[z+d/dz],\;
A^\ast=2^{-1/2}[z-d/dz],\;
\Omega=e^{-z^2/2}\; \pi^{-1/4}
\ee
be the (real valued) 
Hermite ONB of $L_2(\mathbb{R},dz)$ for which we have consequently
the completeness relation
\be \label{2.8}
\delta(z-z')=\sum_{n=0}^\infty \; e_n(z)\; e_n(z')
\ee
Then combining (\ref{2.6}), (\ref{2.7}) we find for $x,x'>0$
\be \label{2.9}
\delta(x-x')=(1+\frac{1}{x\;x'})\sum_{n=0}^\infty\; e_n(f(x))\; e_n(f(x'))
\ee
For $x,x'<0$ we find analogously unique $z,z'$ such that 
$x=h_-(z),\;x'=h_-(z')$. Then analogously
\ba \label{2.10}
\delta(x-x') &=& \delta(h_-(z)-h_-(z'))=
\frac{\delta(z-z')}{|dh_-/dz(z')|}
\nonumber\\
&=& \delta(z-z')\;\frac{2w(z')}{|h_-(z')|}
=-\delta(z-z')\;\frac{2w(z')}{h_-(z')}
=-\delta(z-z')\;\frac{h_+(z')-h_-(z')}{h_-(z')}
\nonumber\\
&=& \delta(z-z')\;[1+\frac{1}{h^2_-(z')}]
= \delta(z-z')\;[1+\frac{1}{h_-(z)\;h_-(z')}]
\ea
so that (\ref{2.9}) is also valid for $x,x'<0$. Finally for 
$h_-(z)=x<0<x'=h_+(z')$ ($x'<0<x$ analogous) we have 
\be \label{2.11}
0=\delta(x-x')=(1+\frac{1}{h_+(z)\; h_-(z')})\delta(z-z')
\ee
due to (\ref{2.3}) so that (\ref{2.9}) continues to hold also in this 
case. Thus we have shown
\begin{Proposition} \label{pro2.1} ~\\
The functions 
\be \label{2.12}
g_{0,n}(x):=e_n(f(x)),\; g_{1,n}(x):=\frac{1}{x}\;e_n(f(x));\;\;
n\in \mathbb{N}_0
\ee
provide an orthonormal basis of $\cal H$, that is 
\be \label{2.13}
<g_{j,m},\; g_{k,n}>=\delta_{j,k}\;\delta_{m,n},\;
\sum_{j=0,1}\;\sum_{m\in \mathbb{N}_0}\; g_{j,m}\;<g_{j,m},.>=1_{{\cal H}}
\ee
\end{Proposition}
For $a\not=1$ we just have to substitute $f(x)$ by $\sqrt{a}f(x)$ and 
multiply $g_{j,k}$ by $\sqrt{a}$.\\
\\
Since $e^{-[x^2+x^{-2}]/2}=e^{-1/2}\;e^{-z^2/2}\propto e_0$ and $e_m$ is 
an even/odd polynomial in $z$ of degree $m$ times $e_0$ for $m$ 
even/odd, the $g_{j,m}$ lie 
in the finite linear span of the $b_n,\, n\in \mathbb{Z}$. To see that 
the operators $Q,Q^{-1},P$ (multiplication by $x,x^{-1}$ or derivation by 
$x$ respectively) preserve the finite linear span of the $g_{j,m}$ we note 
\ba \label{2.14}
x\; g_{0,m}(x) &=& [z+\frac{1}{x}]g_{0,m}(x)=g_{1,m}(x)+
2^{-1/2} ([A+A^\ast]\;e_m)(f(x))          
\nonumber\\
x^{-1}\; g_{0,m}(x) &=& g_{1,m}(x)
\nonumber\\
x\; g_{1,m}(x)&=& g_{0,m}(x)
\nonumber\\
x^{-1}\; g_{1,m}(x) &=&[-z+x]\;g_{1,m}(x)=g_{0,m}(x)-x^{-1}
2^{-1/2} ([A+A^\ast]\;e_m)(f(x))          
\nonumber\\
\frac{d}{dx} g_{0,m}(x)&=&
\frac{dz}{dx}\; \frac{d}{dz} e_m(z)
=(1+x^{-2})\;2^{-1/2} ([A-A^\ast]\;e_m)(f(x))
\nonumber\\
\frac{d}{dx} g_{1,m}(x) &=&
-x^{-2} \; g_{0,m}(x)+x^{-1} \frac{d}{dx} g_{0,m}(x)
\ea
Since $A e_m=\sqrt{m}\; e_{m-1},\;A^\ast e_m=\sqrt{m+1} e_{m+1}$ it follows
that above operators preserve the finite linear span of the ONB.

We also note that $g_{0,m}$ is a linear combination of the $b_n,\;
|n|\le m$ while $g_{1,m}$ is a linear combination of the $b_n,\;
-(m+1)\le n\le m-1$. Hence the sequence 
$g_{0,0}, g_{1,0},\; g_{0,1}, g_{1,1}, g_{0,2}, g_{1,2},..., g_{0,m}, 
g_{1,m},..$ is 
obtained by Gram-Schmidt
orthonormalisation of the seqence of vectors 
$b_0, b_{-1}, b_1,\; b_{-2}, b_2, b_{-3},.., b_m, b_{-(m+1)},..$ 
in precisely 
this order. 

We verify by elementary means the orthonormality of the system $g_{j,k}$
\ba \label{2.15}
<g_{j,m},\;g_{k,n}> 
&=& [1+(-1)^{j+k+m+n})]\;
\int_0^\infty\;dx\; x^{-[j+k]}\;e_m(f(x))\;e_n(f(x)) 
\nonumber\\
&=& [1+(-1)^{j+k+m+n})]\;
\int_{-\infty}^\infty\;dz\; 
\frac{h_+(z)}{2 w(z)}\; h_+(z)^{-[j+k]}\;e_m(z)\;e_n(z) 
\ea
In the first line we have split the integral over the whole 
real axis into the integral over the negative real axis and 
used that $f(-x)=-f(x),\; e_n(-z)=(-1)^n\; e_n(z), (-x)^j=(-1)^j x^j$.
In the second we changed integration variables from $x$ to $z$.
If $j+k=0$ then (\ref{2.15}) is non-vanishing only if $m,n$ are both 
even or odd, thus $e_m e_n$ is even and thus the term $z/2$ in $h_+$ drops 
out and the integral collapses to $\delta_{m,n}$. 
If $j+k=2$ then (\ref{2.15}) is non-vanishing only if $m,n$ are both 
even or odd, thus $e_m e_n$ is even and thus the term $-z/2$ in 
$h_+^{-1}=-h_-$ drops 
out and the integral collapses to $\delta_{m,n}$. 
If $j+k=1$ then (\ref{2.15}) is non-vanishing only if $m,n$ are not both 
even or odd, thus $e_m e_n$ is odd and $w^{-1}$ is even, hence the integral 
vanishes.

\section{Applications for the positive real line}
\label{s3}

Applications for $x$ having range only in $\mathbb{R}^+$ naturally arise 
when the potential in question has spherical symmetry so that $x=r$ is a 
radial variable. 
Then the Hilbert space measure is $r^{D-1}\; dr$ if 
one works in $D$ spatial dimensions. 
Or for some reason we may be interested not in the operators $x,x^{-1}$ but in 
$|x|,\;|x|^{-1}$. We may still use the basis 
$b_n(x)$ as a smooth invariant domain but now the inner products are 
no longer computable analytically for all applications. However, we show 
in this section how to estimate them. 

Consider for $n\in \mathbb{Z}$
\be \label{3.1}
J_n=\int_0^\infty\; dx \; x^n \; e^{-[x^2+x^{-2}]/2}
\ee
In contrast to \cite{1}, $n$ is not constrained to be even because we 
integrate only over the positive real line. However, like in \cite{1} we see 
that $J_n=J_{2-n}$ by a change of variables $x\mapsto x^{-1}$ which 
now means that we need to compute $J_n$ only for $n>0$ rather than $n=2m\ge 0$.
Using the methods of \cite{1} we can compute $J_n$ analytically for 
$n$ even. We provide here an alternative way to see this. 
\ba \label{3.2}     
J_n &=& e^1\;
\int_{-\infty}^\infty\; \frac{dz}{2w} \; [h_+]^{n+1}\; e^{-z^2/2} 
\nonumber\\
&=& \int_{-\infty}^\infty\; \frac{dz}{2w} \; 
\sum_{k=0}^{n+1}\; 
\left( \begin{array}{c}
n+1 \\ k
\end{array}
\right) (z/2)^k\; w^{n+1-k} \; e^{-z^2/2} 
\nonumber\\
&=& \int_{-\infty}^\infty\; \frac{dz}{2} \; 
\sum_{k=0}^{[(n+1)/2]}\; 
\left( \begin{array}{c}
n+1 \\ 2k
\end{array}
\right) (z/2)^{2k}\; w^{n-2k} \; e^{-z^2/2} 
\ea
which shows that for $n$ even we just have to compute the Gaussian integral 
of $z^{2k},\; k=0,..,n/2$ (here $[.]$ denotes the Gauss bracket). 
For $n$ odd this is no longer analytically possible due to the square 
root $w$. But we can use the basic 
estimate 
\be \label{3.3}
w^{-1}<1<w
\ee
to show that for $n$ odd 
\ba \label{3.4}
&& \int_{-\infty}^\infty\; \frac{dz}{2} \; 
\sum_{k=0}^{(n-1)/2}\; 
\left( \begin{array}{c}
n+1 \\ 2k
\end{array}
\right) (z/2)^{2k}\; w^{n-1-2k} \; e^{-z^2/2} 
\nonumber\\
&<&  J_n 
\nonumber\\
&<&
\int_{-\infty}^\infty\; \frac{dz}{2} \; 
\sum_{k=0}^{(n+1)/2}\; 
\left( \begin{array}{c}
n+1 \\ 2k
\end{array}
\right) (z/2)^{2k}\; w^{n+1-2k} \; e^{-z^2/2} 
\ea
Both upper and lower bound are explicitly computable.

\section{Conclusion and outlook}
\label{s4}

That the functions $b_n$ come with such a high degree of analytical control 
with respect to their Hilbert space applications is quite surprising. 
One would have thought that they are even harder to handle than  
function systems based on $x^n \exp(-p(x)),\; n\in \mathbb{N}_0$
where $p$ is an even polynomial such as $x^2+x^4$ which certainly is 
of relevance for anharmonic polynomial potentials \cite{2}. That one can 
even orthonormalise them in closed form is even more astonishing. The 
mechanism at work here is the simple fact that $x^2+x^{-2}=(x-1/x)^2+2$.
This shows that the $b_n$ are Gaussians in $z=x-1/x$ which therefore explains 
the appearance of Hermite functions in $z$. 

Among the many physical applications that come to mind are Schr\"odinger 
type operators $H$ with an attractive potential decaying at infinity. 
Then the spectrum of $H$ will be of mixed type with discrete part 
corresponding to bound states (true eigenvectors) 
and continuous part corresponding to scattering states (generalised 
eigenvectors). Let ${\cal H}={\cal H}_b\oplus{\cal H}_b^\perp$ be the 
decomposition of the Hilbert space into bound states and the orthogonal 
complement of their closed span. The ONB $g_{j,k}$ and its corresponding 
resolution of 
unity allows for a decomposition of the true eigenvectors in ${\cal H}_b$
and vectors in the complement into the $g_{j,n}$ 
which may prove useful in many applications in which the effect of 
scattering states (more precisely normalised wave packets formed from them) 
and bound states must be considered simultaneously. 

We close by relating the fundamental inner product integral (formula (2.10) in 
reference \cite{1}) 
\be \label{4.1}
I_n(a)=\int_{\mathbb{R}}\; dx \;
\; x^{2n} \; e^{-\frac{a}{2}\;[x^2+x^{-2}]},\;n\ge 0 
\ee
to the modiefied Bessel functions of the second kind 
\cite{3} (see formula 10.32.9)
\be \label{4.2}
K_c(b)=\int_{\mathbb{R}_+}\; dx\; e^{-b\;{\rm ch}(x)}\; {\rm ch}(c\;x)
\ee
namely 
\be \label{4.3}
I_n(a) = 2 K_{n+ 1/2}(a)
\ee
To see this we reduce the integral (\ref{4.1}) to twice its restriction 
to the positive real axis and then substitute
$x=\exp(1/2 z)$ with z in entire real axis. 
Then the integrand becomes 
$\exp((n+1/2)z)\; \exp(-a[\exp(z)+\exp(-z)]/2)$. Then decomposing the $z$ 
integral into positive and negative real axis yields directly (\ref{4.3}). 
Thus (\ref{4.1}) is related to a spherical Bessel function as $n+1/2$ 
is half integral for which altenative closed formulas exist.\\
\\
\\
{\bf Acknowledgements}\\
\\
T.T. thanks Daniel Burgarth and Jorma Louko for fruitful discussions.

\end{document}